# A Review of Machine Learning and Algorithmic Methods for Protein Phosphorylation Sites Prediction

Farzaneh Esmaili[1, #], Mahdi Pourmirzaei[1, #], Shahin Ramazi[2, *], Seyedehsamaneh Shojaeilangari[3], Elham Yavari[1]

{f.esmaili, m.poormirzaie, s.ramazi, e.yavari}@modares.ac.ir, s.shojaie@irost.ir

[#] Equal contribution. [*] Corresponding author

Tarbiat Modares University

Iranian Research Organization (IROST)

## Abstract

Post-translational modifications (PTMs) have key roles in extending the functional diversity of proteins and as a result, regulating diverse cellular processes in prokaryotic and eukaryotic organisms. Phosphorylation modification is a vital PTM that occurs in most proteins and plays a significant role in many biological processes. Disorders in the phosphorylation process lead to multiple diseases including neurological disorders and cancers. The purpose of this review paper is to organize this body of knowledge associated with phosphorylation site (p-site) prediction to facilitate future research in this field. At first, we comprehensively reviewed all related databases and introduced all steps regarding dataset creation, data preprocessing and method evaluation in p-site prediction. Next, we investigated p-sites prediction methods which fall into two computational groups: Algorithmic and Machine Learning (ML). Additionally, it was shown that there are basically two main approaches for p-sites prediction by ML: conventional and End-to-End deep learning methods, which were given an overview for both of them. Moreover, this study introduced the most important feature extraction techniques which have mostly been used in p-site prediction. Finally, we created three test sets from new proteins related to the 2022th released version of the dbPTM database based on general and human species. Evaluation of the available online tools on the test sets showed quite poor performance for p-sites prediction.

# 1 Introduction

PTMs are biochemical reactions occurring on a protein after its translation [1,2] which change the regulated physicochemical properties, maturity, and activity of most proteins [3,4]. PTMs include cutting, folding, ligand-binding, adding a modifying group to one or more amino acids, and finally changing the chemical nature of amino acids [5,6]. In recent years, an increasing volume of PTM data is available because of improvements in mass spectrometry (MS) based on high-throughput proteomics [7]. There are more than 600 types of PTMs [8] that affect many aspects of cellular functionalities, such as metabolism, signal transduction, activity, stability, and localization of various proteins [9,10]. Recent studies have shown that each modification leads to a multitude of effects on the structure and therefore, the function of the proteins [11]. PTMs include phosphorylation, glycosylation, ubiquitination, sumoylation, acetylation, succinylation, nitrosylation as well as numerous others in most cellular activities [9,12–16]. Moreover, PTMs play key roles in a variety of biological regulatory pathways, including like metabolic pathways, DNA damage response, transcriptional regulation, signaling pathways, protein-protein interactions, apoptosis, cell death, insulin signaling, immune response, and aging [17,18]. Dysregulation in PTMs is contributed to cancer, diabetes, cardiovascular disease, and neurological disorder [19–24].

Phosphorylation is one of the most important reversible PTMs. Phoebus Levene discovered phosphorylation in 1906 in the protein vitellin (Phosvitin) [25]. In phosphorylation, a −2 phosphate group is covalently added to Serine (S), Threonine (T), Tyrosine (Y), and Histidine residues and then removed by protein phosphatases. It is known that protein phosphorylation regulates the activity of various enzymes and receptors including signal pathways [26] and can greatly impact the folding, function, stability, and subcellular localization of the protein [25,27,28]. In eukaryotes, this modification plays a vital role in signal transduction and other biological functions protein synthesis, cell division, signal transduction, DNA repair, environmental stress response, regulation of transcription, apoptosis, cellular motility, immune response, metabolism, cell growth, development, cellular differentiation, and aging [29,30].

In eukaryotes, the phosphorylation process is catalyzed via Protein Kinases (PKs) differentially and specifically which each PK only modified a subset of substrates to ensure signaling fidelity [31]. Phosphorylation is present in more than one-third of human proteins, and this modification is regulated by approximately 568 human PKs and 156 protein phosphatases [29]. In this sense, phosphorylation is one of the widest spread and most extensively studied protein PTMs which  has a significant role in the control of biological processes like proliferation, differentiation, and apoptosis [29,32]. Site mutations or dysregulation of kinase activity, their hyperactivity, malfunction or overexpression and also, hyper phosphorylation of human proteins are associated with certain disease states such as cancers, Alzheimer’s disease (AD), Parkinson's disease (PD), frontotemporal dementia (FTD), and various pathways involving the immune system [27–29,33]. Therefore, identifying kinase-specific p-sites is essential for understanding the regulatory mechanisms of phosphorylation.

Multiple experimental methods are used for the detection assays of protein phosphorylation like liquid chromatography-tandem mass spectrometry (LC-MS/MS), radioactive chemical labeling, and immunological detection, such as proximity ligation assay (PLA), chromatin immunoprecipitation, and western blotting. Although, the combination of LC-MS/MS method with IP strategy is suitable for detection of p-sites in proteins [34,35], however, the use of experimental approaches is very expensive and challenging to detect p-sites on a large scale. This cause those computational methods for distinguishing PTMs have been appealing to scientists. Cellular processes are regulated by phosphorylation, which is highly conserved and affects protein stability in a significant way. **Figure 1** shows the protein phosphorylation scheme.

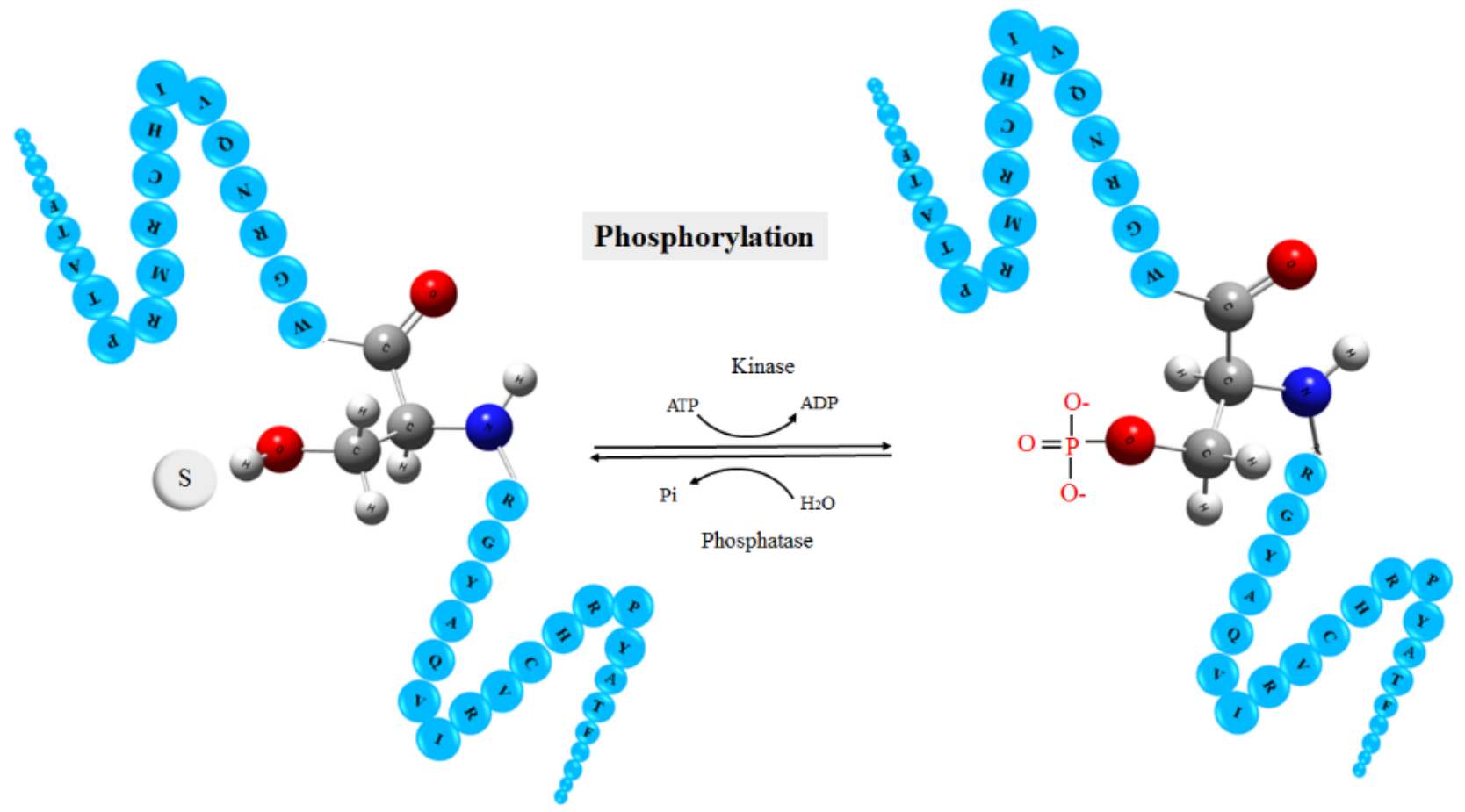

**Figure 1:** Schema of protein phosphorylation [34].

The number of known phosphorylated sites has grown since 2003, it rose from 2,000 to more than 500,000 known sites in experimental databases. P-sites are involved in the regulation of at least 30% of human proteomes [36,37].

Indeed, experimental approaches are generally difficult, slow, costly, and need specialized equipment and knowledge. Over the last two decades, PTM research has made remarkable progress due to technological advancements and the emergence of new computational methods.

A study [34] reviewed PTM tools, resources, and related databases and also they investigated the challenges of algorithmic methods. Ramazi et al. [34], divided 10 types of PTMs into small chemical groups, lipids, and small proteins. They investigated databases and algorithmic approaches for different PTM sites. Shi et al [38], reviewed 19 available tools for phosphorylation networks. They reported different analyses for their functionality, data sources, performance, network visualization, and implementation. Rashid MM et al. [39], reviewed specified ML methods, main feature selection methods, databases, and current online tools for microbial p-sites. They only investigated microbial p-sites and did not mention other p-sites in organisms nonetheless. Also, their work was limited to classical ML methods.

In this study, unlike other previous studies, we investigate all features, databases, and methods concerning p-sites prediction. The contribution of this work is summarized as follows:

- Valid PTMs databases that contain phosphorylation experimental data were introduced. Then, the two most important phosphorylation databases were reviewed in which the number of organisms and p-sites were covered in detail.
- Two main data preparing p-sites datasets steps include data collection and data preprocessing were reviewed. In other words, this study investigated methods for data collection and also introduced the most important and functional approaches for data preprocessing. Additionally, all evaluation metrics which have been used for p-sites prediction were introduced.
- Most common and important feature extraction methods in four types of structural level, sequential, evolutionary, and physicochemical property-based were described.
- It was found that there are generally two machine-learning-based approaches exist for p-sites prediction which we divided into conventional ML and End-to-End deep learning methods. In the present study, methods of both approaches were reviewed and available online tools of p-site prediction were briefly introduced.

- In the end, we created three test sets from new proteins related to the 2022nd released version of the dbPTM database, and then evaluated and compared the available online tools together in different metrics on the three specific test sets.

## 2 Databases

Developing a prediction model requires a dataset of experimentally validated phosphorylation. Therefore, the availability of general and specific databases for p-sites is the first step toward this end [34]. Databases are constantly evolving due to the advent of technology and may be updated by providing accurate details. These databases contain information of different organisms such as viruses, animals, and plants that has been collected manually and experimentally. For instance, all the information in HPRD has been collected manually where it contains more than 95,000 phosphorylation sites extracted from ~13,000 proteins [40].

Considering different types of PTMs, databases are arranged into specific and general terms, where general PTM databases cover a wide domain of data for different types of PTM, but specific databases are constructed based on special types of PTMs like phosphorylation.

Databases such as dbPTM [41], SysPTM [3], SwissProt [42] and HPRD [40] are general databases which cover different types of PTMs and p-site is one of them. On the other hand, EPSD [43], Lymphos2 [44], phospho3d [45], phosphoELM [46] and Regphos [47] are specifically gathered for p-sites.

In the following, two important databases, dbPTM and EPSD, which are known as general and specific databases for p-sites data are going to introduce. Furthermore, **Table 1** summarize both general and specific databases according to their statistical information for p-sites.

### 2.1 dbPTM

Database post-translation modification (dbPTM) is a general database that integrates PTM's data from 30 databases and ~92,600 research articles. The dbPTM covers 130 types of PTMs in more than 1,000 organisms [34]. The 2022th version of dbPTM [48] has curated more than 2,777,000 PTM sites from 41 published databases and ~82,000 research articles.

### 2.2 EPSD

Eukaryotic Phosphorylation Site Database (EPSD) is one of the most specific and comprehensive databases for p-sites which has been updated in 2020. EPSD has been updated from two databases of dbPPT [49] and dbPAF [50], which includes roughly ~82,000 p-sites for 20 plants and more than 483,000 p-sites from seven different types of animals and fungi. Moreover, EPSD collected p-sites from 13 additional databases including PhosphoSitePlus [51], Phospho.ELM [52], UniProt [53], PhosphoPep [54], BioGRID [55], dbPTM, FPD [56], HPRD, MPPD [57], P3DB [58], PHOSIDA [59], PhosPhAt [60], and SysPTM [36]. Totally, this database contains ~1,616,800 experimentally known p-sites in approximately 209,300 phosphoproteins of 68 eukaryotes (18 animals, 24 plants, 19 fungi, and 7 protists). **Figure 2, 3, and 4** depict the EPSD database p-sites details.

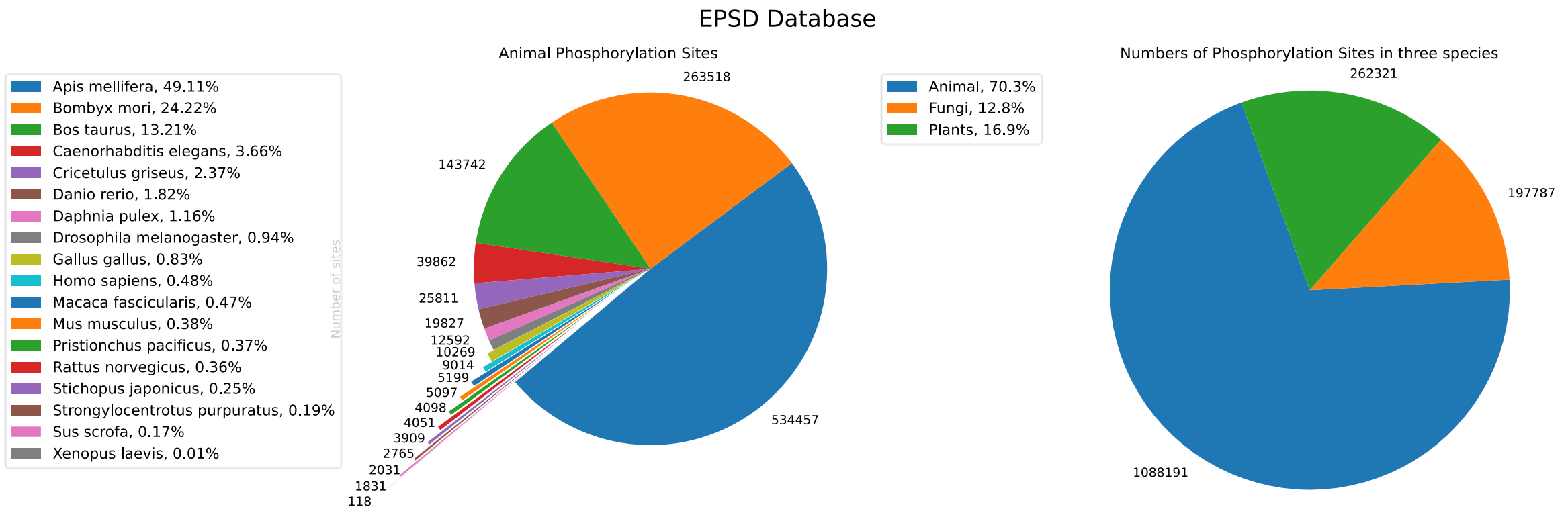

**Figure 2: (Left)** shows the number of phosphorylation sites (p-sites) in the animal proteins distributed by different types of animals and **(right)** shows the number of p-sites on proteins related to three species: animal, fungi and plants. All figures are based on EPSD database.

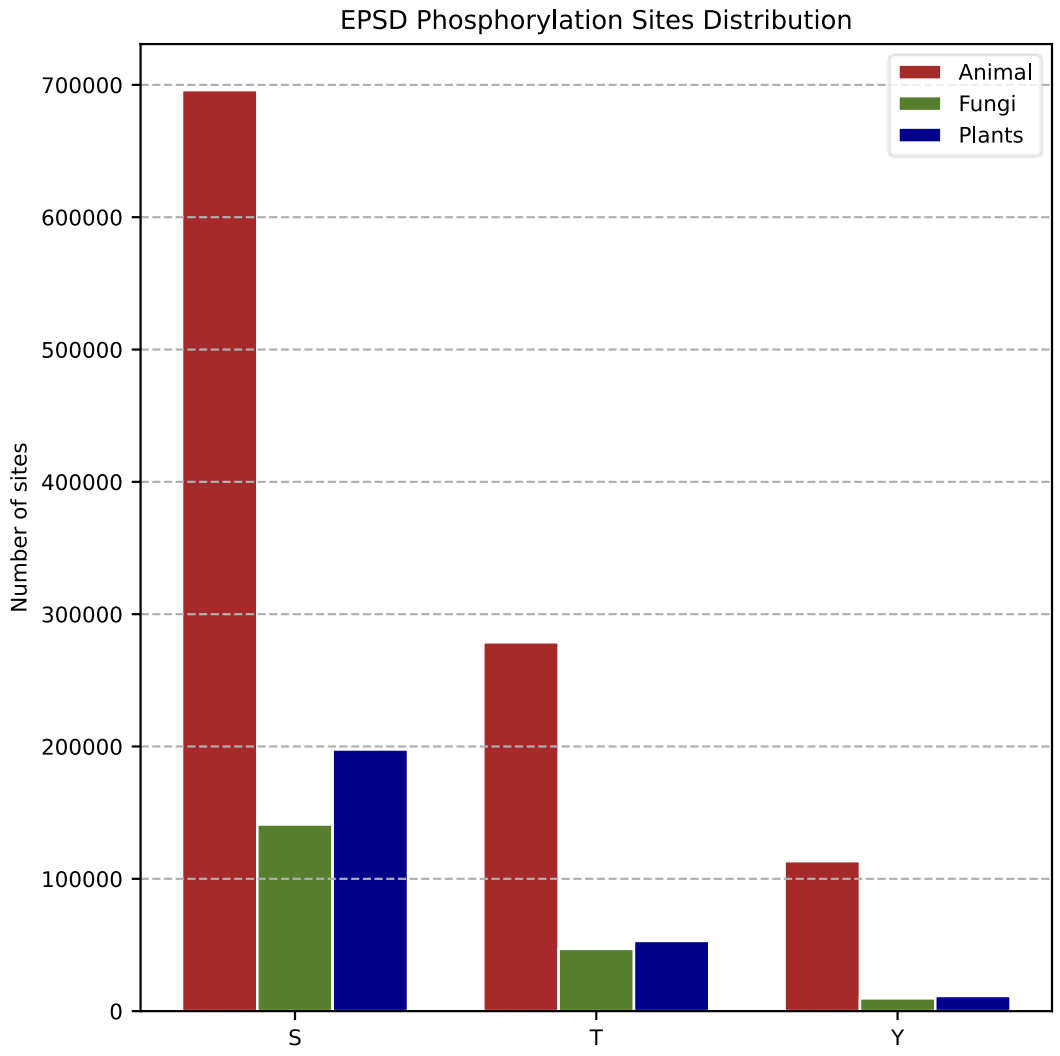

**Figure 3:** Number of S, T, and Y in proteins related to animal, plants, and fungi organisms in EPSD database.

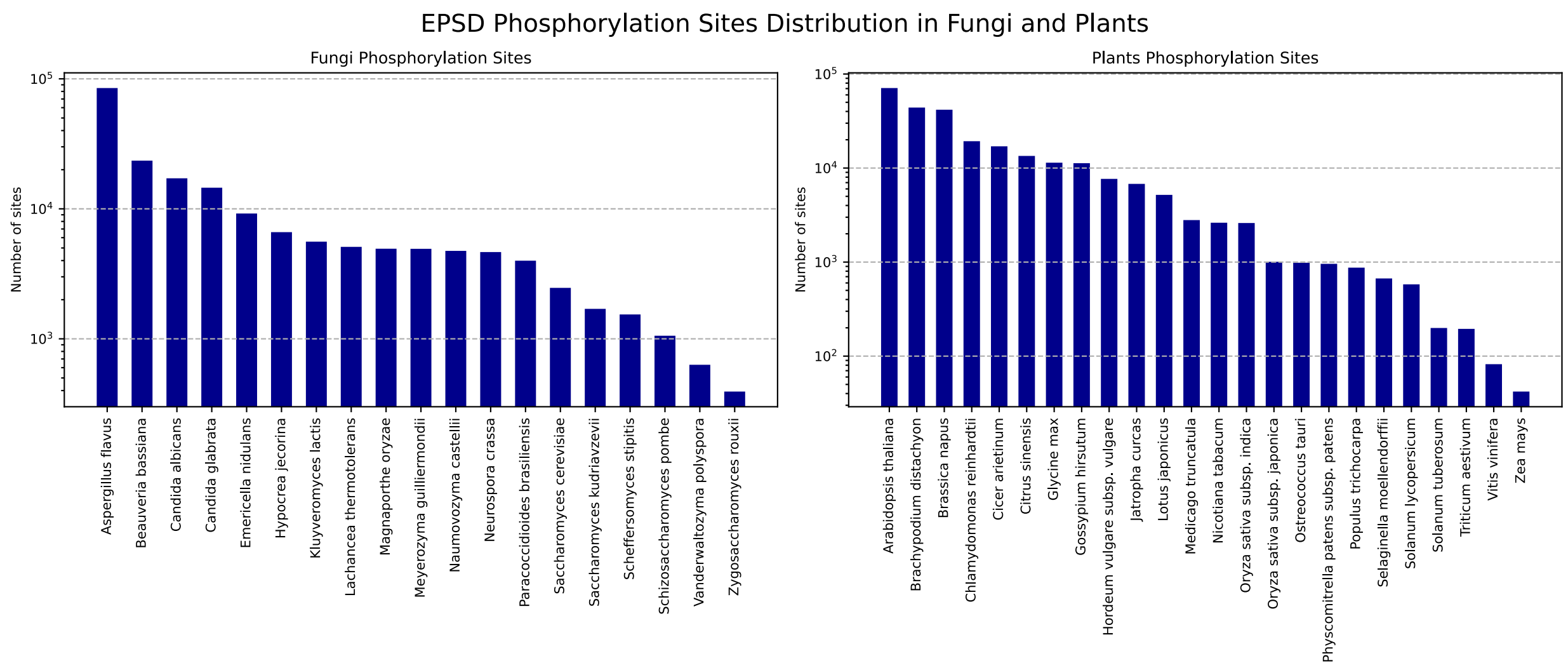

**Figure 4:** P-sites distribution in EPSD database in log scale for **(left)** fungi proteins and **(right)** plants proteins.

Table 1: Contains two general and specific databases which cover number of p-sites and proteins (P). Moreover, it provides useful information about each one. This table is inspired by [34]. * Type of database can be secondary or primary; secondary databases are the integration of other databases. ** Primary databases are independent.

| **Type** | **Acronym** | **General statistics** | | **Type of data and database** | **URL** |
|---|---|---|---|---|---|
| | | **Number of covered organisms** | **Number of phosphorylation sites** | | |
| **General database** | dbPTM [41] | More than 1,000 organisms | p-sites: ~1,770,000 P: ~557,700 | Experimental and Predicted Secondary* | https://awi.cuhk.edu.cn/dbPTM/ |
| | Phosphosite Plus [61] | 26 organisms | p-sites: ~240,000 P: ~20,200 | Experimental Primary | https://www.phosphosite.org |
| | PTMCode v2 [62] | 19 organisms | p-sites: ~316,500 P: ~45,300 | Experimental Secondary | http://ptmcode.embl.de |
| | qPTM [63] | Human | p-sites: ~199,000 P: ~18,402 | Experimental Secondary | http://qptm.omicsbio.info/ |
| | YAAM [64] | Saccharomyces cerevisiae | p-sites: ~3,900 P: ~680 | Experimental Secondary | http://yaam.ifc.unam.mx |
| | HPRD [40] | Human | p-sites: ~1,100 P: ~30,000 | Experimental Primary | http://www.hprd.org |
| | PHOSIDA [59] | 9 organisms | p-sites: ~70,000 P: ~28,700 | Experimental Secondary | http://www.phosida.com |
| | PTM-SD [1] | 7 model organisms | p-sites: ~1,600 P: ~842 | Experimental Secondary | http://www.dsimb.inserm.fr/dsimb_tools/PTM-SD |
| | SysPTM [3] | 6 organisms | p-sites: ~353,000 P: ~53,200 | Experimental Secondary | http://lifecenter.sgst.cn/SysPTM/ |
| **Phosphorylation databases** | EPSD [43] | 68 organisms | p-sites:~1,616,800 P: ~209,300 | Experimental Secondary | http://epsd.biocuckoo.cn |
| | PhosphoNET [65] | Human | p-sites: ~966,000 P: ~20,000 | Experimental and Predicted | http://www.phosphonet.ca |
| | | | | Secondary | |
| | RegPhos [47] | Human, mouse and rat | p-sites: ~113,000<br>P: ~18,700 | Experimental and Predicted Secondary | http://140.138.144.141/~RegPhos |
| | Phospho.ELM [46] | Mainly model organims | p-sites: ~42,500<br>P: ~8,600 | Experimental Secondary | http://phospho.elm.eu.org |
| | Phospho3D [45] | Mainly model organisms | p-sites: ~42,500<br>P: ~8,700 | Experimental Secondary | http://www.phospho3d.org |
| | dbPSP [66] | 200 prokaryotic organisms | p-sites: ~19,300<br>P: ~8,600 | Experimental Secondary | http://dbpsp.biocuckoo.cn/indExp.php |
| | pTestis [67] | Mouse | p-sites: ~17,800<br>P: ~3,900 | Experimental and Predicted Secondary | http://ptestis.biocuckoo.org |
| | LymPHOS [44] | Human Mouse | p-sites: ~18,300<br>P: ~4,900 | Experimental and Predicted Primary | https://www.lymphos.org |
| | P3DB [58] | 45 plant organisms | p-sites: ~220,000<br>P: ~57,000 | Experimental and Predicted | http://www.p3db.org |

## 2.3 Identifying driver mutations and their effects on p-site proteins

Phosphorylation is involved in many aspects of cellular organization and signaling pathways associated with the disease. Various studies have demonstrated that p-sites are evolutionarily constrained in human genomes, as well as prevalent in cancer driver mutations and causal variants of inherited disease. Therefore, phosphorylation information and knowledge of its function are useful for interpreting genetic variation, genotype-phenotype associations, and molecular disease and their treatment [68].

DNA single nucleotide variants (SNVs) are caused by a single nucleotide change, which is the most common type of sequence changes. Genetic variation of p-sites via SVNs can directly have an effect on modifying target residues or indirectly by modifying the consensus binding sequences (i.e., short linear motifs) located in the flanking sequences of phosphorylated residues. As a result, this can change signaling networks by making, changing, and disrupting the p-sites [69]. There have been reports of phosphorylation-related SNVs that disrupt existing sites and create new sites, disturb the kinase-substrate interactions and cause disease phenotypes. A major challenge facing biomedical research is the identification of genotype phenotype associations, molecular mechanisms, and cancer driver mutations [68].

There are various databases with a useful list of genome variants in p-sites and other PTM sites. However, they provide no perspective of how mutations on p-sites and other protein sites will affect kinase binding. Therefore, databases and updated tools are required to interpret rapidly increasing genomic and phosphoproteomic data to explain the signaling networks. We are briefly going to describe the ActiveDriverDB database as well as MIMP and PTMsnp tools in this field.

The ActiveDriverDB is a web database which was designed to understand how protein coding varies in the human genomes. The ActiveDriverDB database contains more than 260,000 experimentally identified PTM sites in the human proteome using public databases like PhosphoSitePlus, UniProt, Phospho.ELM, and HPRD which contains ~149,300 p-sites. As evidenced in the ActiveDriverDB database, changes in target amino acids substitutions in p- sites influence the creation of pathogenic disease mutations, somatic mutations in cancer genomes, and germline variants in humans. Additionally, the ActiveDriverDB database contains phosphoproteomics data reflecting the cellular response to SARSCoV-2 infection, which can be used to predict the impact of human genetic variation on COVID-19 infection and disease course [69].

Mutation impact on phosphorylation (MIMP) (http://mimp.baderlab.org/) is an online tool for predicting kinase-substrate interactions based on missense SNVs. MIMP analyzes kinase sequence specificities and predicts whether SNVs disrupt the existing p-sites or create the new ones. This helps discover mutations that modify protein function by altering kinase networks and provides insight into disease biology and therapy development [68].

PTMsnp is another online tool for identifying driver genetic mutations aiming at PTM sites in proteins across different cohorts from TCGA by using a Bayesian hierarchical model. There are more than 411,500 modification sites in PTMsnp from 33 different types of PTMs and 1,776,800 mutation sites from 33 types of cancer. The web server detects proteins with a higher frequency of PTM-specific mutations in the motif region, considered to be the key targets in human disease development [70].

## 3 Data gathering and preprocessing

In this section, we are going to describe steps concerning creating and preprocessing datasets before p-sites prediction. In the last decade, due to the importance of phosphorylation in understanding biological systems of proteins and in guiding basic biomedical drug design, research on phosphorylation has boomed. Several experimental methods are used to identify p-sites in a large number of phosphorylation examples with high accuracy but many of them are labor-intensive and time-consuming. Therefore, low-cost and fast algorithmic and ML methods have become popular to overcome the problems associated with experimental methods [71]. In order to build a dataset for p-sites prediction, all verified data from multiple databases are considered. Mainly, there are two main steps to prepare  a dataset [34,71].

1. Data collection
2. Data preprocessing

**Negative data collection**: S, T, and Y amino acids existing in experimental peptides without any phospho-groups are considered as non-p-sites or negative samples.

There are two major strategies available to choose the negative samples:

- From phosphoproteins, the negative random samples of the target residue that did not undergo the phosphorylation modifications are selected.
- From non-phosphoproteins with none of their target residues (S, T, and Y) that have undergone specific phosphorylation, (based on experimental evidence) are selected as the negative set.

**Positive data collection**: S, T, and Y amino acids as p-sites or the positive samples are usually compiled from the aforementioned databases (e.g., EPSD and dbPTM). These samples are usually have known from experimental experiments.

### 3.1 Data preprocessing

After constructing the primary positive and negative datasets, one important task is removing inconsistent/redundant samples to gain a more reliable dataset.

Cluster Database at High Identity with Tolerance (CD-HIT) program is a protein clustering program widely used to reduce the sequence homology and filter out the similar ones. According to different phosphorylation prediction studies [71–74], a threshold of identity is considered to range from 30% to 60% in many phosphorylation prediction studies [75].

Three main policies in the literature for removing inconsistent/redundant proteins are as follows  [34,76]:

1. Removing redundant phosphoproteins using the CD-HIT program.
2. Removing identical subsequences within the positive and negative sets by choosing the size of the optimal windows.

3. Removing identical subsequences between the positive and negative datasets by choosing the size of the optimal windows.

## 3.2 Class imbalanced problem

It is a common problem in ML when there is an imbalance between the distribution ratios of data classes. In other words, a dataset that has unequal samples in classes is imbalanced. This is not an issue when the difference is not that much. Nevertheless, when one or more classes are infrequent, many models do not work well at identifying the minority classes. For example, in p-site prediction, preprocessed datasets are mostly imbalanced because the number of the negative samples is much greater than positive samples. **Figure 5** shows the preprocessing framework for balancing data.

In the following, three most used approaches to deal with class imbalanced problems are introduced:

**Up-sampling**: It generates additional data for minority classes either by making copies of the minimum class or by creating synthetic data which can represent samples of minimum classes**.**

**Down-sampling**: It removes data from the majority class either randomly or using intelligent approaches of sample selection to handle the issue.

**Customized loss function**: This is a technique to deal with imbalance problems in ML that tries to customize the model loss function by assigning larger weights to minority. Customized losses demonstrated better performance and have attracted more attention than up sampling and down sampling approaches [77].

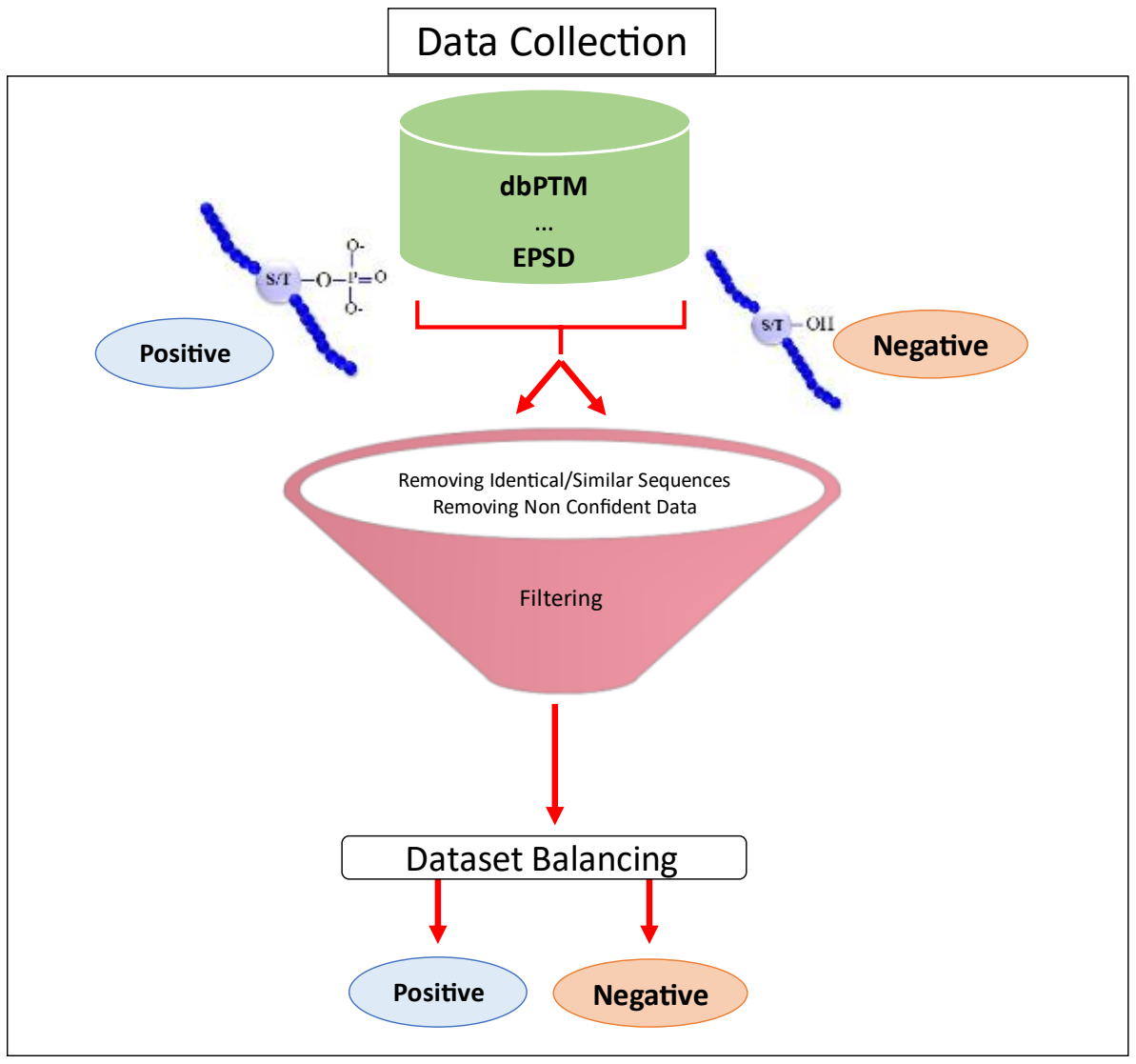

**Figure 5:** Data preprocessing flow which includes the balancing step.

## 4 Evaluation

The well-known evaluation metrics for protein p-sites are classified into five methods: Accuracy (ACC), Sensitivity (SN), Specificity (SP), Matthews Coefficients of Correlation (MCC), and the Area Under the ROC Curve (AUC). These metrics are evaluated with a confusion matrix that summarizes the performance of models; it compares the real target values with those predicted by a model. The number of rows and columns in this matrix is based on the number of classes. From the confusion matrix, we will end up with four values [34,76]:

**True positive (TP)**: Represents the number of positive samples classified correctly.
**False Positive (FP):** Represents the number of negative samples classified incorrectly.
**True Negative (TN):** Represents the number of negative samples classified correctly.
**False Negative (FN):** Represents the number of positive samples classified incorrectly.

ACC is the percentage of correct predictions. This metric is defined in **Equation 1** as the ratio of correctly classified samples (both TP and TN) to the total number of cases examined.

$$Accuracy = \frac{TP + TN}{TP + TN + FP + FN} \tag{1}$$

SN or Recall represents the ratio of TP prediction to the total number of TP and FN (**Equation 2**).

$$Recall = \frac{TP}{TP + FN} \tag{2}$$

SP is defined in **Equation 3**. It computes the ratio of samples predicted truly to the sum of the TN and FP.

$$Specificity = \frac{TN}{TN + FP} \tag{3}$$

The Precision metric is defined in **Equation 4**. It represents the ratio of TP samples to all cases predicted positive.

$$Precision = \frac{TP}{TP + FP}$$

(4)

The F1-score or F-Measure is a combination of Precision and Recall which sums up the predictive performance of the model. The formula is showed in **Equation 5**.

$$F1 = \frac{2 \times Precision \times Recall}{Precision + Recall}$$

(5)

Two SN and SP measures are used to plot the ROC curve, and AUC is used to determine the model performance. Furthermore, if the problem is a binary classification, it is informative to report the MCC metric (**Equation 6**) which indicates by increasing the correlation between true and predicted values, the prediction will become better.

$$MCC = \frac{TP \times TN - FP \times FN}{\sqrt{(TP + FN)(TP + FP)(TN + FN)(TN + FP)}}$$

(6)

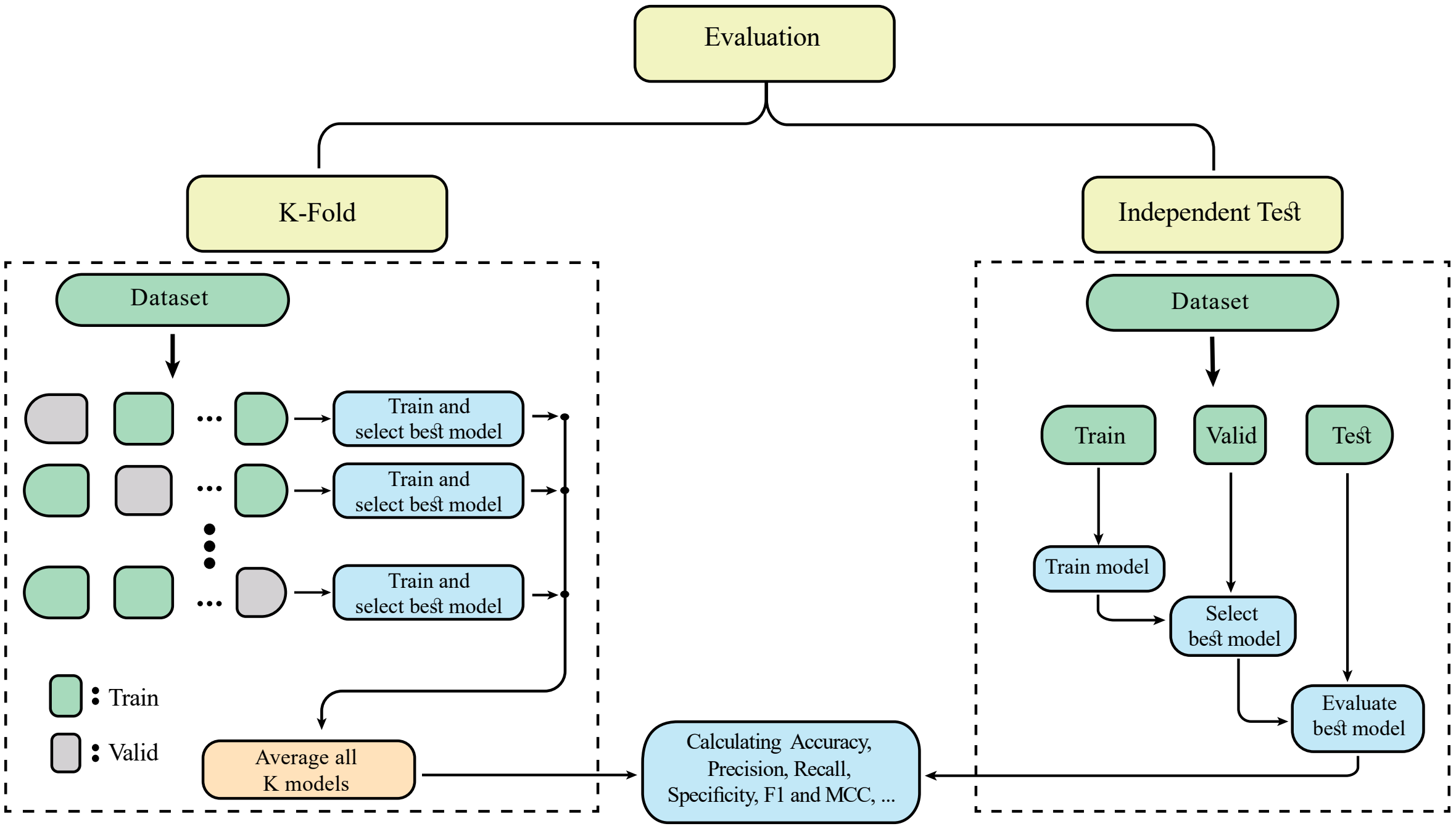

**Figure 6:** The evaluation step can be done by two methods: K-fold cross validation and independent test. Independent test method sometimes is called "Train-Test" or "Train-Valid-Test" as well.

### 4.1 Model evaluation

Basically, there are three methods for model evaluation for p-site prediction: Independent test (Train-Test) and K-fold cross-validation and Jack-Knife cross validation (or the leave-one-out cross validation). In the first one, a dataset is split into two sets: a train and a test set. Then, the train set is divided into two subsets again: a train set and a valid set. The basic procedure is that the train set is used to train models and the valid set is used for the evaluation of the trained models. After selecting the best model with respect to the valid set result, we need to evaluate it on the test set. At the end, we should report the test set and there shouldn't be much difference between the valid set and the test set results (**Figure 6**).

On the other hand, there is another assessment strategy utilized to assess ML models on restricted data sample. The method contains a single parameter called k that alludes to the number of bunches that data samples should be divided into. That is why the procedure is called k-fold cross-validation which specific values for k can be chosen. Considering the scenario of 5-Fold cross-validation (k=5), a dataset is divided into 5 bunches. Within the first iteration, the primary fold is utilized to assess the model and the rest are utilized to train the model. Within the second iteration, subsequent fold is utilized as the validation set whereas the rest serve as the training set. This process is repeated until each fold has been used as the validation set. Each sample is given the opportunity to be utilized within the validation set one time as well as utilized to train the model k-1 times. The k-fold Cross-Validation is usually used when the amount of Train-Valid data is limited. On the contrary, when dealing with huge amounts of data, we do not need to have a big valid set. In other words, the proportion of the Train-Valid split sometimes can go below 1% for the valid set. This approach is mostly used when massive amounts of data are accessible. But in low data regimes, they usually split with proportions of 70%-30%.

Note that there is also another evaluation strategy named Jack-knife validation test [78] rarely used for p-site prediction. As the most objective method, the jack-knife cross validation (or the leave-one-out cross validation) delivers unique results for a dataset in which one sample is selected to serve as the test data, while the rest are used as the training data. This procedure is repeated N times for a dataset with N samples, which can be expensive for large data sets [79].

In summary, K-fold should be used in low data regimes and an independent method with a small percentage of a test set should be used when we have access to lots of data.

## 5 Methods for predicting phosphorylation sites

In the following sections, we are going to review methods of p-sites classification by dividing them into two main categories: Algorithmic methods and ML. Likewise, ML methods are also divided into two approaches: conventional ML and End-to-End deep learning methods. These two are going to be described in the following.

### 5.1 Algorithmic methods

Innovative algorithms based on statistical approaches have been used in many studies. Here, we need to define algorithmic methods as computational methods in which there are no learning algorithms to gain information directly from data. Schwartz and Gygi [80], proposed a statistically repetitive method, using a set of phosphorylated peptide sequences to extract the patterns and a set of peptide sequences to evaluate the predictions. They mapped two sets of sequences to the position-weight matrix so that in the matrices, the number of repetitions of each residue was determined from 6 positions higher to 6 positions lower than each p-site (it means their window size for each peptide is 13 amino acids long). Then, they formed a binary matrix based on these two matrices. This final matrix indicates the probability of observing a specific residue around a p-sites by examining this matrix and comparing it with other p-sites.

Chen et al. [81] presented a new method for predicting p-sites by collecting four background datasets including phosphorylated and non-phosphorylated sequences. They chose a given length of 13 for windows around p-sites. Initially, they formed the weight-position matrices and then, extracted the patterns. By scoring those patterns and deleting some of them, they finally reported a series of patterns as the output during an iterative cycle.

He et al. [82], showed that the number of patterns to be examined around each positions are growing exponentially based on the length of the window. They refer to two developed algorithms to find phosphorylation patterns, named the Motif-X and the MoDL algorithms. They supposed that these algorithms do not detect all patterns and some patterns remain hidden from biologists. Therefore, they introduced a new algorithm called Motif-ALL to discover and report all possible patterns based on previous algorithms.

There has been a family of algorithms called Group-based Prediction System (GPS) for many years as algorithmic methods [83–89]. In 2004, an algorithm was developed, group-based p-site predicting and scoring 1.0, based on the hypothesis that similar short peptides exhibit similar biological functions. Likewise, the algorithm was refined and created an online service of GPS 1.1, which could predict p-sites for 71 PK clusters. Then, GPS 2.0 and 2.1 were presented with the same scoring strategy using two methods named matrix mutation (MaM) and motif length selection (MLS) which were designed to improve the accuracy. Consequently, GPS 2.2, 3.0, 4.0, and 5.0 algorithms were developed which are used for the prediction of other PTM sites rather than p-site [31].

### 5.2 Machine learning methods

Most algorithms used for phosphorylation prediction are based on ML. Moreover, with explosions of the DL method in the early 2010s, ML gets popular even more than before. ML is generally the ability of machines to do actions based on prior knowledge and experience [90]. There are more than 40 different methods for predicting p-sites in which many of them is based on ML techniques including Logistic Regression (LR), Support Machine Vector (SVM), Random Forest (RF), and K-Nearest Neighbor (KNN) [71].

In general, there are two main strategies in ML to predict Phosphorylation: conventional ML and End-to-End deep learning methods. The conventional approach stands for using ML algorithms as a part of solving a solution besides other steps in a pipeline design such as feature extraction and hand-feature engineering. In other words, usually, in a conventional ML based system, there has been multiple stages of processing which need to be designed individually. While, the End-to-End deep learning approaches can replace all those steps with a single neural network. This type of learning tries to eliminate the need of explicit feature engineering steps inside the learning system by feeding the raw data as the input to it.

#### 5.2.1 Feature extraction

In protein phosphorylation prediction, various types of conventional approaches have been studied. Feature extraction is an important step of those approaches [91]. In this paper, we reviewed 20 feature extraction techniques suggested according to physicochemical, sequential, evolutionary and structural properties of amino acids. We have tried to introduce the most important and practical methods of feature extraction in the following.

##### 5.2.1.1 Physicochemical property-based features

**Encoding based on grouped weight ( EBGW):** EBGW divides 20 amino acid into 7 categories based on their hydrophobicity and charge characteristics [92,93]. For each group $H_i$ (i =1, 2, 3), a 25-dimensional array $S_i$ (i = 1, 2, 3) of the same element in the segment should be generated. If the amino acid at that position has belonged to the $H_i$ group, the element in the array will set to 1, otherwise, it will set to 0. Each array will be divided into sub-arrays (*J*-ones), which represent as *D(j)*. This value can be taken from cutting the main $S_i$ from the first window with *len (D(j))* defined as **Equation 7**:

$$len\big(D(j)\big) = int\ \left(\frac{j*L}{J}\right)\ j = 1, 2, \dots, J \quad ,\ L = length\ of\ segments \tag{7}$$

For each group of $H_i$, a vector with length of *J* based on its sub arrays should be defined in which the *j*-th element of $X_i^{(j)}$, is calculated based on **Equation 8**:

$$X_i^{(j)} = \frac{Sum\ (D(j))}{len\ (D(j))} \tag{8}$$

**Amino Acid Index (AAINDEX):** The features based on amino acid indices are extracted from AAINDEX database. This database is used to represent various physicochemical and biochemical properties of each amino acid alone and also pairs of them in every PTM [94]. The feature encodes 14 properties: hydrophobicity, polarity, polarizability, solvent /hydration potential, accessibility reduction ratio, net charge index of side chains, molecular weight, PK-N, PK-C, melting point, optical rotation, entropy of formation, heat capacity, and absolute entropy [92,95].

**Average Accumulated Hydrophobicity (ACH)**: ACH quantifies the tendency of amino acids surrounding S, T, or Y residues to be exposed to solvent [96]. For different window sizes, ACH is calculated by averaging the cumulative hydrophobicity indices around the p-site. Note that every site is located in the center of the sliding windows [97,98].

**Encoding scheme Based on Attribute Grouping (EBAG)**: EBAG represents the hydrophobicity attribute of the amino acids and divides the residues into 4 classes based on their physicochemical property: hydrophobic class $c_1$ = {A, F, G, I, L, M, P, V, W}, polar class $c_2$ = {C, N, Q, S, T, Y}, acidic class $c_3$ = {D, E}, and basic class $c_4$ = {H, K, R} [99,100].

**Overlapping Properties (OP)**: OP clusters each protein based on their chemical attributes. Each amino acid is classified into 10 physicochemical properties: polar, positive, negative, charged, hydrophobic, aliphatic, aromatic, small, tiny and proline [98].

**Pseudo amino acid compositions (PseAAC):** This feature is firstly defined by Chou and et.al [101] for coding proteins. They proposed sequence order and physicochemical information in protein sequences. For more details refer to [102–105].

#### 5.2.1.2 Sequence-based features

**Quasi-sequence order (QSO):** It describes the physicochemical distance between amino acids [92]. Most physicochemical properties are hydrophobicity, hydrophilicity, polarity, and side-chain volume. This feature was originally proposed by Chou and et.al [101]. For more detail refer to [101,106].

**Numerical representation for amino acids:** It converts each character of amino acids into numerical numbers by mapping them in alphabetic order from 1 to 20 and dummy amino acid X represents 21 [92].

**Binary encoding of amino acids (BINA):** BINA represents each amino acid as 21-dimensional binary vectors, which encodes 1 for the target amino acid and 0 for the residues (other 20 amino acids). For example, alanine (‘A’) is shown as 100000000000000000000 [92].

**LOGO**: This feature is defined by calculating the occurrence of amino acid frequencies and encoding them in a sequence with Two the Sample Logo program [92].

**Position Weight Amino Acid composition (PWAA)**: Position information of each amino acid is another key point that shall be considered in feature extraction. PWAA can reveal sequence order information around P, S, and Y residues [107]. PWAA can be declared from **Equation 9** which $L$ represents number of upstream or downstream of amino acids from p-sites in specific windows, if $x_{i,j}$= 1 means that each amino acid is belonged to *j-th* position in window, otherwise $x_{i,j}$ = 0.

$$C_i = \frac{1}{L(L+1)} \sum_{j=-L}^{L} x_{i,j} \left( j + \frac{|j|}{L} \right), \quad j = -L, \dots, L \tag{9}$$

**Composition of K-Spaced Amino Acid Pairs (CKAAP)**: The encoding of CKSAAP is pretty easy, which can directly be calculated from the sequence pieces of p-sites and non-p-sites. CKSAAP is one of the important feature encoding schemes in lots of prediction tasks, especially in representing short sequence residues in protein sequence or subsequence. All 21 amino acids contain 441 different possible pairs. For scanning pieces to count all pairs of amino acid with k-space, we can use different window sizes. For example, window AXXV is a two space amino acids pair in $k$ = 2 [73,108]**.** CKSAAP equation is proposed as **Equation 10** [73]. In this equation, $L$ denotes length of window, $A_iA_j$ is amino acids pairs.

$$f_{i,j} = \frac{Num\ (A_iA_j)}{L - K - 1}\ i, j = 1,2, \ldots 21 \tag{10}$$

**Amino acid compositions (AAC)**: AAC is the most common used feature, which simply calculates each amino acid's frequency in subsequences of a protein while encoding the information into 20 bits [109].This feature is also represented as Amino acid Frequency (AF) in some researches. Both AF and AAC reflect frequency of each amino acids or amino acid pairs occurrence. Lin et al. [109] proposed AAC equation as **Equation 11** which $c_i$ is number of amino acid *i* in the sequence and $v_i$ refers to AAC.

$$v_i = \frac{c_i}{len(seq)}\ \ i = 1, \ldots., 20 \tag{11}$$

### 5.2.1.3 Evolutionary-based features

**K-Nearest Neighbor (KNN):** The most popular feature selection method which is used in various ML problems, especially in PTM and phosphorylation classification is KNN. It classifies sequences based on their distance. The algorithm classifies sequences by looking at k of nearest neighbor sequences by finding out the majority votes from nearest neighbors that have similar attributes and the shortest distance as those used to map the items [110].

**Position-Specific Scoring Matrix-based transformation (PSSM):** PSSM encodes the evolutionary data of a protein which is very informative and useful for some biological classifications problems. The PSSM matrix in a protein with a sequence of length L is a matrix with L * 20 dimensions. In the matrix, each row represents an amino acid in the protein sequence, and the columns represent the 20 amino acids in proteins [111].

### 5.2.1.4 Structural-based features

**Protein Disorder Features** (**DF**): All PTMs include p-sites located within disorder positions [112]. Protein disorders were used as features in many studies [97,113,114].

**Shannon Entropy (H)**: Entropy in information theory quantifies the amount of uncertainty of a random variable. To be more precise, it is the average (Expected value) amount of information obtained from observing a random variable. It means, when the entropy of a random variable is high, we have more ambiguity about that random variable [115]. In science and engineering in general, entropy is a measure of the degree of ambiguity or disorders [116].

**Relative Entropy (RE)**: It is known as Kullback Leibler which is aggregated entropies for more than 20 sites in proteins[74].

**Information gain (IG):** IG can be computed by subtracting RE from entropy (**Equation 12**) [98].

$$IG = H - RE \tag{12}$$

**Accessible Surface Area (ASA)**: ASA or solvent-accessible surface area is a biomolecule surface which can access the solvent. This is an essential structural feature determining the proteins folding and stability[98].

### 5.2.2 Conventional machine learning approach

Once the features have been extracted, classification models should be adopted to predict the p-sites. One of the most popular classifier is SVM [97,109,117].

SVM is a linear model for classification and regression problems which uses a line or hyperplane to separate data. In other words, SVMs calculate maximum margin boundary that leads to equivalent division of all data points. First, SVM uses a line to classify each data point based on their distance. If data points are not linearly separable in low dimensional space, there may be multiple transformations enabling the data be linearly separable in higher dimensions. Therefore, SVMs can find a hyperplane in higher dimensions between different classes of data such that the distance between data points falling on either side of that hyperplane to be maximized [118,119]. Nowadays, SVMs have been widely used in bioinformatics, especially in PTM problems [97,107,120].

RF is another well-known and important classifier frequently used in this field. This algorithm can randomly build a forest which contains a large number of decision trees. Each tree constructs a class prediction and the class with the maximum votes will become the model prediction [121]. **Figure 7** demonstrates the procedure of feature extraction and also, **Figure 8** shows the process of conventional ML methods.

As of recent years, kinases-specific methods have been used since in general, some protein prediction sites have not yet been explored and kinases can assist in locating these sites. NetPhos [122] and NetPhosK [123] both used DNN based on consensus sequences and mass spectrometry experimental methods. These algorithms are specific to the kinase's family. In the Quokka framework [30]**,** LR approach was suggested to classify 43 S/T and 22 Y kinases family sites. Kim et al. [117] proposed to use the consensus sequence structure as features and SVM classifier to predict four kinases groups and families. The best accuracies achieved by their model were reported around 83 - 95% at the kinase family level, and 76–91% at the kinase groups. Liu **et al.** [124] proposed a method for prediction of four kinases families based on RF which extracts features with Auto Covariance (AC) transform and seven physicochemical properties and achieved over 90% accuracy.

To recognize protein p-sites in universal proteins, Huang et al. [107] proposed a method based on SVM in viruses. They used EBAG and PWAA features for extracting physicochemical and sequence information of viral proteins around p-sites. They used 10-fold cross-validation and independent test set for different window sizes from 15-27 lengths. They got the best results for window size 23 with an accuracy of 88.8%, 95.2%, and 97.1% for S, T and Y sites respectively. They also showed the influence of using different features. Their model improved almost 15% when they used the combination of two EBAG and PWAA features.

Furthermore, Lin et al. [109] used KNN, AF, and CKAAP as features and combined different features together to feed it to their model to investigate the best features. The combination of AF and CKAAP provided the best accuracy for their SVM model. They believed SVM could classify rice protein as universal p-sites. Their work was named Rice Phospho 1.0 which achieved 82% accuracy.

Cheng et al. proposed Granularity SVM(GSVM) for predicting universal p-sites [97]**.** They used KNN, AF, DF, and ACH features in every p-sites position for making the train set. To split data into high-dimensional feature spaces, they used kernel fuzzy c-means clustering which utilized it as feature extraction method. The method was applied to plants and animal dataset types and could achieve 80% and 85% accuracy, respectively.

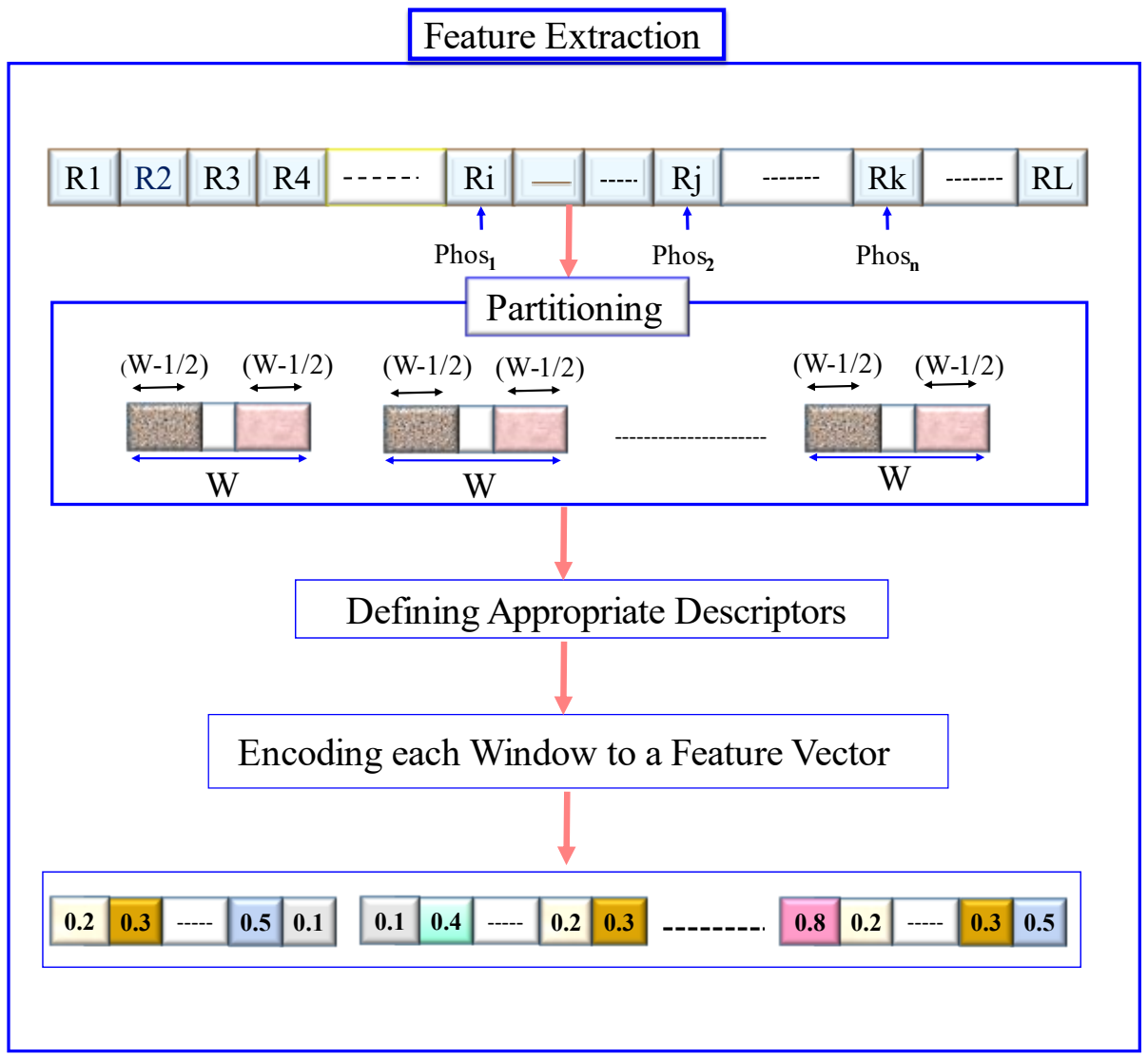

**Figure 7:** A common procedure in feature extraction stage.

By Phospred-RF method, Banerjee et al. [125] used information extracted from PSSW and trained individuals RF with odd window sizes from 9-25 amino acids. They got approximately 70% accuracy for 26 protein sequences. RF-phos-1.0 transformed each amino acid to vectors by using eight algorithms of feature selection (H, RE, ASA, OP, ACH, ACC, QSO, and the sequence order coupling number of each sequence) based on 9 amino acid windows size. They specifically showed which features are the most important and have more effects on accuracy. It was mentioned that AAC was the best feature for S and T sites. Then, these features were used as RF input with 10-fold-cross validation. The accuracy of the model was approximately 80% for S, T, and Y sites [74]. Moreover, in the RF-phos-2.0, their RF model was improved by using window sizes of 5 to 21 amino acids and using different features. QSO was the best feature for S and T [98]. RF-phos-1.0 and RF-phos-2.0 specifically predicted universal p-sites. It should be mentioned that feature selection methods helped to improve the accuracy of various approaches.

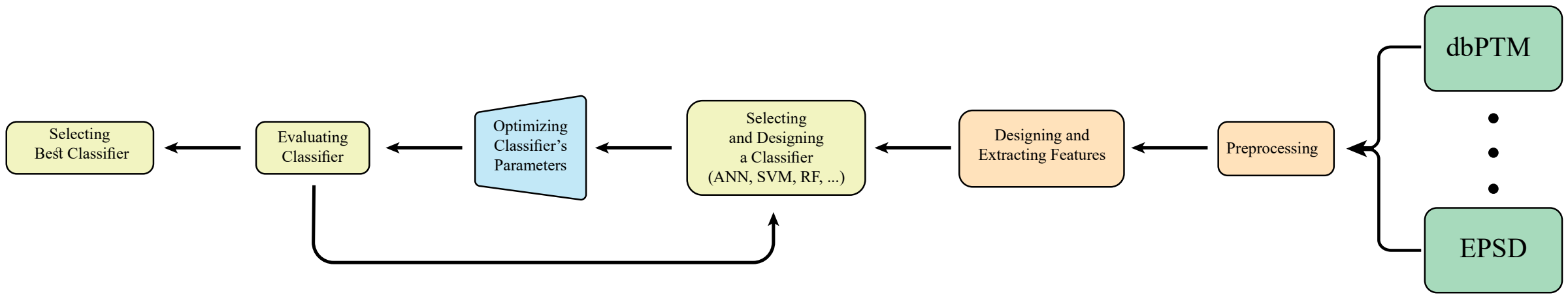

**Figure 8:** Procedure of using conventional Machine Learning (ML) methods.

Microbial Phosphorylation Site predictor (MPsites) was proposed by Hassan et al. [126]to recognize universal microbial p-sites with different sequence features. In order to convert each sequence to numerical vectors, they used various sequence encoding strategies including AF, BE, AAINDEX, and PWAA. They used naïve bayes, SVM, neural networks, decision tree, and RF algorithms to recognize S and T p-sites. Results showed that RF has better performance than the other algorithms. It got 68% accuracy for S sites and 75% accuracy for T sites [126].

Cao et al. [127] proposed a method to predict p-sites in 7 species specific fungi proteins. They used a strategy including two steps for feature optimization to improve the SVM prediction performance. KNN, AAC, di-Amino Acid Composition and physiochemical properties were used as features. First with RF model, they sorted each input features based on the mean accuracy. In the second step, top ten features from previous step were merged to train the SVM model. Finally, they achieved over 80% accuracy.

Chen et al. [128] proposed a feature selection method named GAS, based on ant colony and genetic algorithm for classification of six kinases types.

Qui et al. [105] developed an approach called iPhos-PseEvo. Protein sequence evolutionary and pseudo amino acid composition (PseAAC) were selected as features for ensemble an RF model. The accuracy for their model was 71% with the jackknife test evaluation approach.

As a final example in this section, Multi-iPPseEvo [104] is similar to iPhos-PseEvo but with a different implementation strategy while using k-fold cross validation. This method contains a multi-ensemble RF classifier for each S, T, and Y site and proposed multi-label p-site prediction for each site.

### 5.2.3 End-to-End deep learning approach

End-to-End learning has become a hot topic in ML field by taking the advantage of DL. DNN is almost the same as traditional Artificial Neural Networks (ANNs), which is the composition of many connected neurons that work together to solve specific issues, inspired by the functionality of biological neural networks in the human brain. Inspired by the human brain, each DNN's layer (or group of layers) could be used for learning the hierarchical abstraction for downstream tasks. In other words, usually raw input sequences are just fed to a DNN and the process of feature selection automatically happens between layers. Since it refers to training a possibly complex learning system by applying gradient-based learning to the system as a whole, it is called End-to-End deep learning. These systems are specially designed so that all components are created to be differentiable and consequently, be learnable. That is to say, it is a procedure in which a model learns all the steps including feature selection and extraction between the first and the last layers [129]. **Figure 9** shows the common procedure of End-to-End deep learning methods.

Need to mention that in order to prepare a sequence of amino acids for the End-to-End deep learning system, there are two prerequisite steps [130]: 1) sequence encoding, 2) converting the encoded sequence to numerical vectors. The second step could be done via either one-hot encoding [72] or another popular technique named word embedding [131]. Therefore, one-hot encoding is not considered as a feature extraction method and is simply used to represent categorical inputs (e.g., amino acids codes) into numerical vectors in order to feed to DL models. However, in an End-to-End deep learning network, word embedding is often used in PTM due to the similarity between PTM and Natural Language Processing (NLP) domains as well as the effectiveness of the technique.

We have shown a great success of DL in solving problems in different domains of science, especially in biological problems with finding non-obvious patterns or making prediction in datasets [131–137]. In recent years, DL has been applied to PTM classification of proteins such as p-site prediction. As mentioned earlier, the main aspect of this approach compared to the conventional ML approach is that the feature extraction step is not designed by human engineers or manually. These layers are acquired from input data to extract best patterns accurately and quickly. Though, the most important point about DL is that it needs huge amounts of data and in fact, by increasing the size of dataset, it can perform better. This can be counted as a drawback; when the dataset is not big enough, it quickly falls behind other ML methods in terms of performance.

Among all DL architectures, Convolutional Neural Networks (CNN), Recurrent Neural Networks (RNN) and Long Short Memory (LSTM) are the famous ones [71,72,138].

Wang et al. [72] provided a DL architecture called MusiteDeep to predict general and kinases-specific families' position in a sequence. The window size used for input sequence was 33. Then, they presented their network with multi-layer CNN and attention layers architecture. In DeepPhos' paper [71], in contrast to multi-layer models of MusiteDeep, it was decided to use dense CNN blocks that could show different and multiple representations of proteins for p-sites predictions by using the concatenation of intra block layers and inter block layers. The method could improve the performance of MusiteDeep by using different window sizes with length of 15, 33, and 51. Both of these methods DeepPhos and MusiteDeep have been developed for kinases family and universal p-sites. Moreover, PhosTransfer [138] is a DL based framework which constructed a pre-train architecture with CNNs based on hierarchy kinases system and transfer learning. It was specialized for improving kinases p-sites prediction. The method was to accumulate the information of a hierarchical kinase's classification tree at family, subfamily and group levels. It could achieve 0.89 AUC scores on average. The DeepPPSite is another DL model based on universal p-sites prediction with considering sequence information [73]. Ahmed et al. used one hot encoding sequence as input, PSPM, EBGW, CKSAAP, and AAINDEX as features and stacked LSTM architecture as a predictive model. The MCC value reported for S, T and Y is 0.358, 0.356 and 0.350, respectively.

Furthermore, there have been some researches such as Lv et al. 's work [139] which used hybrid architectures. They presented a specific hybrid End-to-End architecture that combined both CNN and LSTM together called DeepIPs, to predict universal p-sites in host cells infected with SARS-CoV-2 [140,141]. Lv et al. utilize three approaches in Natural Language Processing as word embedding layers to represent amino acids as vectors: Glove [142], Fast Text [143,144] and Word2vec [130] pre-training word embedding methods. The final accuracy for this method was reported as 80.45 for S/T and 75.22 for Y.

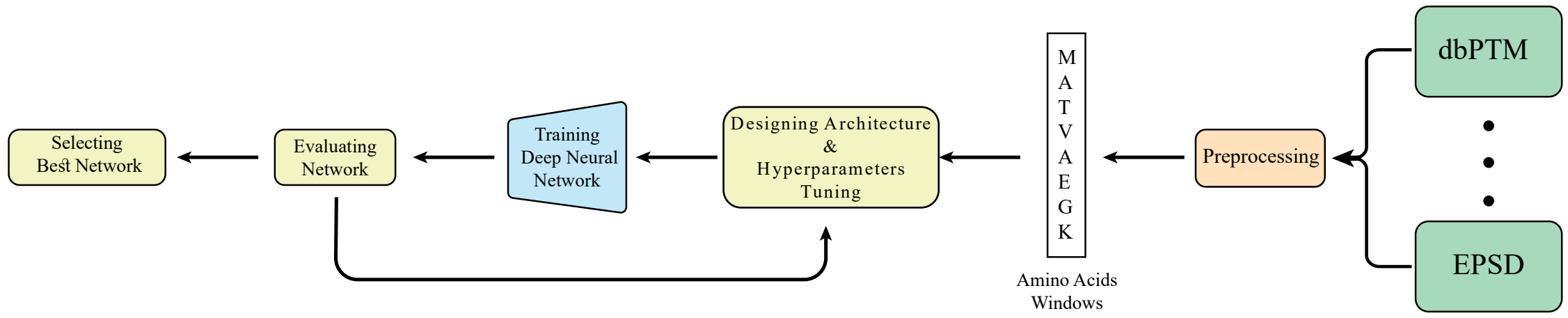

**Figure 9:** Procedure of End-to-End deep learning method.

DL provides a highly effective framework for dealing with modern-day learning challenges. The modern high-performance interpretable deep tabular learning network (TabNet) provides an extremely powerful framework for solving more challenging learning problems [145]. For example, Khalili et al. [76] developed a TabNet model to predict p-sites in soybean with a high accuracy rate that outperformed other common ML methods (LR-L1, LR-L2, RF, SVM, and XGBoost). They assessed and compared the strength and reliability of all models using 10-fold cross-validation. Experiments assessed the performance of AAC, DPC, TPC, PSSM, and physicochemical properties as individual features. To extract training sequences for model development, various window sizes ranging from 7 to 35 values were used. They got the best results for window size 13 with an accuracy of 87.34% based on PSSM features.

Naser et al. [146] compared human-based feature representation with DL based representation for the reorganization of Phosphoserine p-sites. The combination of RNN-LSTM model got 81.1% accuracy and CNN based achieved 78.3 %. In contrast to human engineering with 77% accuracy, DL methods have performed better for Phosphoserine p-sites prediction.

Even though, most DL approaches worked well with large volumes of data, a paper [147] with small amount of data from only two kinases family proposed a simple DNN architecture and achieved around 80% accuracy. It means that end-to-end learning can also perform successfully on low data regions. This algorithm was designed for both kinases family and universal p-sites.

Guo et al. [148] collected phosphoprotein-binding domains (PPBD) that interact with phosphoprotein-binding domains containing proteins (PPCP) from 12 eukaryotic species and developed a transfer learning based DNN framework to classify the protein binding domains into a hierarchical structure with three levels, including group, family, and single PPBD cluster.

Despite most of End-to-End approaches which used the raw sequences (one-hot encoding) as input, PhosIDN [149] trained a DNN with combining raw sequences and protein-protein interaction information together. This architecture contains three sub-networks: a) sequence feature encoding sub-network (SFENet), b) PPI feature encoding sub-network (IFENet), c) heterogeneous feature combination sub-network (HFCNet).

Table 2: models and tools for p-sites proteins prediction. *: U stands for universal which includes all types of p-sties and K stands for kinase which includes only kinase-specific p-sites **: is not available at the time of writing.

| | **Acronym** | **Type / Description** | **Method** | **Feature extraction method** | **Dataset size** | **Window size** | **Negative dataset** | **Unbalance strategy** | **Redundancy threshold** | **Evaluation strategies** | **URL** | **U/K *** |
|---|---|---|---|---|---|---|---|---|---|---|---|---|
| **Phosphorylation** | NetPhos [122] | Conventional | ANN | Sequence composition features | 902 p-sites | 21 (Y, S) 25 (T) | - | - | - | 5-fold | http://www.cbs.dtu.dk/services/NetPhos/ | K |
| | [117] | Conventional | SVM | - | 855 S, 216 T | 3-25 | Phospho-proteins | Down Sample | 70% | 7-fold | http://www.ngri.re.kr/proteo/PredPhospho.html ** | K |
| | [124] | Conventional | RF | Auto covariance transform, 7 physicochemical properties | 1,911 p-sites | - | Phospho-proteins | Down Sample | 40% | 5-fold, Independent test | ------ | K |
| | [107] | Conventional | SVM | EBAG, PWAA | 230 S, 61 T, 14 Y | 23 | Phospho-proteins | Down Sample | - | 10-fold, Independent test | ------ | U |
| | Rice-phospho 1.0 [109] | Conventional | RF | AF, CKSAAP, KNN | 4,220 S, 605 T, 141 Y | 25 | Phospho-proteins | Down Sample | - | 10-fold, Independent test | http://bioinformatics.fafu.edu.cn/rice_phospho1.0 | U |
| | GSVM [97] | Conventional | SVM | KNN, AF, DF, ACH | ~50,000 P-sites | 13 | Phospho-proteins | Down Sample | 30% | 10-fold | ------ | U |
| | RF-phos-1.0 [74] | Conventional | RF | H, RE, ASA, OP, ACH, AAC, QSO | ~28,000 p-sites | 5 to 21 | Phospho-proteins | Down Sample | 30% | 10-fold, Independent test | ------ | U |
| | RF-phos-2.0 [98] | Conventional | RF | H, RE, IG, ASA, OP, ACH, AAC, QSO | ~28,000 p-sites | 5 to 21 | Phospho-proteins | Down Sample | 30% | 10-fold, Independent test | http://bcb.ncat.edu/RFPhos/ ** | U |
| | PhosTransfer [138] | Conventional | CNN | H, RE, DF, OP, ACH | ~10,000 S, ~34,000 T, ~3,000 Y | - | - | Down Sample | 40% | Independent test | https://github.com/yxu132/PhosTransfer | K |
| | deepPsites [73] | Conventional | LSTM | CKSAAP, EBGW, IPCP, PSPM | ~7,000 S, ~2,000 T, ~700 Y | 15, 19, 21 | Phospho-proteins | Down Sample | 30% | 10-fold, Independent test | https://github.com/saeed344/DeepPPSite | U |
| | GPS 5.0 [31] | Conventional | LR | Structural features | ~15,000 p-sites | 20 | Phospho-proteins | Down Sample | - | 10-fold | http://gps.biocuckoo.cn | K |
| | MPSite [126] | Conventional | RF | AF, IP, PSSM, PWAA, SSF | ~2,700 S, 2,100 T | 7 to 25 | Phospho-proteins | Down Sample | 30% | 10-fold, Independent test | http://kurata14.bio.kyutech.ac.jp/MPSite/ | U |
| | Quokka [30] | Conventional | LR | KNN, AF, BLOUSM 64 | ~2,400 S, ~370 T | 15, 19, 21 | Phospho-proteins | Down Sample | 30% | 5-fold, Independent test | http://quokka.erc.monash.edu/#webserver ** | K |
| | PhosContext2vec [150] | Conventional | SVM | H, BLOUSM 64, DF, OP, ACH, Secondary structure | Universal: ~20,000 S, ~5,600 T, ~2,100 Y Kinases: ~4,100 | 25 | Phospho-proteins | Down Sample | - | 10-fold, Independent test | http://phoscontext2vec.erc.monash.edu/ | K/U |
| | PhosphoSVM [120] | Conventionlal | SVM | H, RE, Secondary structure, DF, ASA, OP, KNN, ACH | ~25,000 S, ~7,200 T, ~2,700 Y | 15, 19, 21 | Phospho-proteins | Down Sample | 30% | 10-fold, Independent test | http://sysbio.unl.edu/PhosphoSVM/ | U |
| | PhosPred-RF [151] | Conventional | RF | H, RE, IG, OP | ~4,300 S, ~2,700 T | 15, 19, 21 | Phospho-proteins | Down Sample | 30% | 10-fold, Independent test | http://bioinformatics.ustc.edu.cn/phos_pred/ ** | U |
| | [127] | Conventional | SVM | Sequence information, Evolutionary information, Physicochemical properties | Various for organisms | - | Phospho-proteins | Down Sample | 30% | 10-fold, Independent test | http://computbiol.ncu.edu.cn/PreSSFP ** | U |
| | [128] | Conventional | Multiple classifiers | GAS | ~3,400 p-sites | - | Phospho-proteins | Down Sample | - | 5-fold | ----- | K |
| | iPhos-PseEvo [105] | Conventional | Ensemble - RF | KNN, PseAAC | 845 S, 386 T, 249 Y | - | Phospho-proteins | Down Sample | 50% | Jackknife test | http://www.jci-bioinfo.cn/iPhos-PseEvo ** | U |
| | Multi-iPPseEvo [104] | Conventional | RF | KNN, PseAAC | 845 S, 386 T, 249 Y | - | Phospho-proteins | Down Sample | 50% | 5-fold | http://www.jci-bioinfo.cn/Multi-iPPseEvo ** | U |
| | deepIPs [139] | End-to-End | CNN-LSTM | - | 5,387 S/T, 102 Y | 33 | Phospho-proteins | Down Sample | 30% | Independent test | https://github.com/linDinggroup/DeepIPs. ** http://lin-group.cn/server/DeepIPs/ | U |
| | DeepPhos [71] | End-to-End | CNN | - | 140,000 S/T, 27,000 Y | 15, 33, 51 | Phospho-proteins | Down Sample | 40% | 10-fold, Independent test | https://github.com/USTCHIlab/DeepPhos ** | U/K |
| | MusiteDeep [72] | End-to-End | CNN + attention | - | ~35,000 S/T, ~2,000 Y | 33 | Phospho-proteins | Down Sample | 50% | 5-fold, Independent test | https://www.musite.net/ https://github.com/duolinwang/MusiteDeep_web | U/K |
| | [147] | End-to-End | DNN | - | ~1,800 S, 700 T, 200 Y | 9 | Phospho-proteins | Down Sample | 20% | 10-fold | ------- | U/K |
| | PhosIDN [149] | End-to-End | SFENet + IFENet+ HFCNet | PPI graph embedding | ~160,000 p-sites | 15, 33, 71 | Phospho-proteins | Down Sample | 40% | Independent test | https://github.com/ustchangyuanyang/PhosIDN | U/K |
| | [76] | Neural network + feature | TabNet | AAC, DPC, TPC, PSSM, physicochemical properties | ~ 4 500 p-sites | 7 to 35 | Phospho-proteins | Down Sample | 40% | 10-fold | ------- | U |

### 6. Protein phosphorylation prediction tools

Due to the high cost and low speed of using experimental methods to recognize p-sites, in recent years many computational online tools have been developed to help and increase the quality of p-sites prediction. **Table 2** introduced famous publicly accessible online tools or GitHub repositories for p-sites prediction.

## 7. Current limitations

In general, it is unfair to compare different ML algorithms applied to p-sites prediction task to choose the best technique due to variation in preprocessing steps, evaluation methods, and more importantly database diversity in literatures. Therefore, we tried to evaluate several tools by creating three new test datasets. For this purpose, we selected the 2022 released version of dbPTM [48] database and picked up all new phospho-proteins in all organisms which did not exist in the previous versions. Subsequently, we built following test sets:

**161-all**: 161 new proteins with p-sites were randomly selected from 161 new released organisms' proteins (One protein per organism). This test set consists of 13,403 sites in which 402 of them are p-sites. The maximum and minimum length of sequences are 7,096 and 49 respectively.

**161-humans**: 161 proteins with p-sites were randomly selected from new released homo sapiens' proteins. This test set consists of 7,383 sites in which 714 of them are p-sites. The maximum and minimum length of sequences are 921 and 714 respectively.

**100-top**: 100 new proteins with p-sites were randomly selected from top 10 organisms which have the biggest new protein numbers (Ten proteins per organism). This test set consists of 9,321 sites in which 507 of them are p-sites. The maximum and minimum length of sequences are 3,498 and 102 respectively.

Next, we tried to evaluate several universal p-sites prediction tools introduced in **Table 2** on these datasets. However, there were many hurdles in the evaluation stage. Kim et al. [117], RF-phos-2.0, PhosPred-RF, Cao et al. [127], iPhos-PseEvo, Multi-iPPseEvo were not available. Moreover, Rice-phospho 1.0 and PhosphoSVM only take one sequence as input in order to process and since the process was time consuming, we could not evaluate our three test datasets on them. Furthermore, DeepIPs did not have any response to our request. Finally, we selected three tools MusiteDeep, PhosIDN and NetPhos to evaluate. By the way, NetPhos could not predict sequences with length more than 4,000 amino acids and since the 161-all test set had proteins more than that length, we could not evaluate it. **Table 3** shows the results.

Table 3 Online tools evaluation. *Note:* Evaluating of tools on three 161-all, 161-human and 100-top test sets.

| **Tool** | **MusiteDeep** [72] | | | **PhosIDN** [149] | | | **NetPhos** [122] | | |
|---|---|---|---|---|---|---|---|---|---|
| **Test set** | 161-all | 161-humans | 100-top | 161-all | 161-humans | 100-top | 161-all | 161-humans | 100-top |
| **TP** | 168 | 194 | 150 | 249 | 308 | 297 | - | 447 | 339 |
| **FP** | 1656 | 745 | 1044 | 4356 | 1140 | 2597 | - | 3701 | 5349 |
| **TN** | 11378 | 5927 | 7781 | 8678 | 5532 | 6228 | - | 2971 | 3476 |
| **FN** | 201 | 517 | 346 | 120 | 403 | 199 | - | 264 | 157 |
| **Accuracy (%)** | 86.14 | 82.91 | 85.09 | 66.60 | 79.10 | 70.00 | - | 46.30 | 40.93 |
| **Precision** | 0.09 | 0.21 | 0.13 | 0.05 | 0.21 | 0.1 | - | 0.11 | 0.06 |
| **Recall** | 0.46 | 0.27 | 0.3 | 0.67 | 0.43 | 0.6 | - | 0.63 | 0.68 |
| **F1** | 0.15 | 0.24 | 0.18 | 0.1 | 0.29 | 0.18 | - | 0.18 | 0.11 |
| **Specificity** | 0.87 | 0.89 | 0.88 | 0.67 | 0.83 | 0.71 | - | 0.45 | 0.39 |

As table 3 shows, all three tools performed weakly compared to what they reported on their papers. We interpreted from the results that there are not valid benchmarks for p-sites prediction. In other words, each paper proposed a method applied on a unique test set to report the results, which makes it difficult to compare different methods together. Therefore, for the fair and precise competition, we suggest that the uniform, comprehensive, unique and well-defined test benchmarks for p-sites prediction will be prepared as a crucial step for the future research of this field.

## 8. Conclusion

Almost all proteins contain phosphorylation, which is responsible for critical functions in the cell. Various diseases can be caused by disruptions of this modification. Discovery of phosphorylation as one of the most important PTMs by high-throughput experimental methods is labor-intensive and time-consuming.  Therefore, it is an urgent to develop a tool or method to automatically predict the p-sites. As we investigated the literatures, there is not a complete review paper for p-sites predictions based on ML algorithms. Due to the importance of the issue, this paper briefly introduced some popular PTM databases (including phosphorylation), methods and online tools for p-site prediction to provide a guide to current research.

In this review, we introduced two important databases: EPSD and dbPTM, while compared them in term of p-sites distribution. Then, we gave a brief overview of protein p-sites prediction by ML techniques which are mainly divided into classical ML and End-to-End deep learning methods. In addition to ML, we slightly discussed algorithmic methods as well. Algorithmic methods have statistical basis which are slow and have high time complexity. On the other hand, ML algorithms which are quite popular these days have attracted a lot of attention in p-sites prediction including SVM, LR, and RF. In conventional methods, SVM has shown better performance, although, the feature extraction step would obviously have a significant impact on the final result. Therefore, this study introduced 20 important and widely used feature extraction methods based on the structural, sequential, evolutionary, and physicochemical property-based categories. Additionally, CNN and RNN based architectures known as efficient End-to-End learning styles, were introduced which are able to predict p-sites directly from the raw input sequences without any feature extraction step.

In the next stage, the evaluation methods for predicting p-sites approaches were reported to give the standard metrics for comparison between performances. Finally, in order to demonstrate the current limitation in p-sites prediction methods, we created three test sets and evaluated several available online tools. All those methods performed poorly compared to what was reported in the related papers which shows the importance of creating uniform and well-defined benchmarks for p-sites prediction.

## Acknowledgments

We gratefully acknowledge Hadi Pourmirzaei and Mohammad Ezati from Faculty of Biological Sciences, Tarbiat Modares University, Tehran, Iran for preparing pictures, helps and recommendations.

## Abbreviation List

| | |
|---|---|
| ASA | Accessible Surface Area |
| ACC | Accuracy |
| AD | Alzheimer’s disease |
| AAC | Amino acid compositions |
| AAINDEX | Amino Acid Index |
| AUC | Area Under the ROC Curve |
| ANNs | Artificial Neural Networks |
| AC | Auto Covariance |
| ACH | Average Accumulated Hydrophobicity |
| BINA | Binary encoding of amino acids |
| CD-HIT | Cluster Database at High Identity with Tolerance |
| CKAAP | Composition of K-Spaced Amino Acid Pairs |
| CNN | Convolutional Neural Networks |
| dbPTM | Database post-translation modification |
| EBGW | Encoding based on grouped weight |
| EBAG | Encoding scheme Based on Attribute Grouping |
| EPSD | Eukaryotic Phosphorylation Site Database |
| FN | FALSE Negative |
| FP | FALSE Positive |
| FTD | frontotemporal dementia |
| GSVM | Granularity SVM |
| GPS | Group-based Prediction System |
| HFCNet | heterogeneous feature combination sub-network |
| IG | Information gain |
| KNN | K-Nearest Neighbor |
| LC-MS/MS | liquid chromatography-tandem mass spectrometry |
| LR | Logistic Regression |
| LSTM | Long Short Memory |
| ML | Machine Learning |
| MS | Mass Spectrometry |
| MaM | Matrix Mutation |
| MCC | Matthews Coefficients of Correlation |
| MPsites | Microbial Phosphorylation Site predictor |
| MLS | Motif Length Selection |
| MIMP | Mutation impact on phosphorylation |
| NLP | Natural Language Processing |
| OP | Overlapping Properties |
| PD | Parkinson's disease |
| PPBD | phosphoprotein-binding domains |
| PPCP | phosphoprotein-binding domains containing proteins |
| PWAA | Position Weight Amino Acid composition |
| PSSM | Position-Specific Scoring Matrix-based transformation |
| IFENet | PPI feature encoding sub-network |
| DF | Protein Disorder Features |
| PKs | Protein Kinases |
| PLA | proximity ligation assay |
| PseAAC | Pseudo amino acid compositions |
| QSO | Quasi-sequence order |
| RF | Random Forest |
| RNN | Recurrent Neural Networks |
| RE | Relative Entropy |
| SN | Sensitivity |
| S | Serine |
| H | Shannon Entropy |
| SNVs | single nucleotide variants |
| SP | Specificity |
| SFENet | sequence feature encoding sub-network |
| SVM | Support Machine Vector |
| TabNet | tabular learning network |
| T | Threonine |
| TN | TRUE Negative |
| TP | TRUE positive |
| Y | Tyrosine |